\documentclass[aps,prd,twocolumn,preprintnumbers,amsmath,amssymb,floatfix,nofootinbib]{revtex4}

\usepackage{dcolumn}
\usepackage{bm}
\usepackage[dvips]{graphicx}
\usepackage{amsmath}
\usepackage{epsfig}
\usepackage{amsfonts}
\usepackage{amssymb}

\newcommand{\lsim}{\buildrel < \over {_\sim}}
\newcommand{\gsim}{\buildrel > \over {_\sim}}

\newcommand{\be}{\begin{equation}}
\newcommand{\ee}{\end{equation}}
\newcommand{\bee}{\begin{equation*}}
\newcommand{\eee}{\end{equation*}}
\newcommand{\bea}{\begin{eqnarray}}
\newcommand{\eea}{\end{eqnarray}}
\newcommand{\bean}{\begin{eqnarray*}}
\newcommand{\eean}{\end{eqnarray*}}

\usepackage{color}

\begin{document}

\preprint{ACFI-T13-02}

\title{Probing the Higgs Portal at the LHC\\ Through Resonant di-Higgs Production}

\author{Jose M. No$^{1}$ and Michael Ramsey-Musolf$^{2,3}$}

\affiliation{$^{1}${\it Department of Physics and Astronomy, University of Sussex, 
BN1 9QH Brighton, United Kingdom}
\\
$^{2}${\it Amherst Center for Fundamental Interactions\\
Department of Physics, University of Massachusetts Amherst\\
Amherst, MA 01003 USA}\\
and\\
$^{3}${\it Kellogg Radiation Laboratory, California Institute of Technology\\
Pasadena, CA 91125 USA}}

\date{\today}

\begin{abstract}
We investigate resonant di-Higgs production as a means of probing extended scalar sectors that include a 125 GeV 
Standard Model-like Higgs boson. For concreteness, we consider a gauge singlet Higgs portal scenario {leading} to two mixed doublet-singlet 
states, $h_{1,2}$. For $m_{h_2}> 2 m_{h_1}$, the resonant di-Higgs production process $pp\to h_2\to h_1 h_1$ will lead to  
final states associated with the decaying pair of Standard Model-like Higgs scalars. We focus on $h_2$ production {\em via} gluon fusion and 
on the $b{\bar b}\tau^+\tau^-$ final state. We find that discovery of the $h_2$ at the LHC may be achieved with $\lsim$ 100 fb$^{-1}$ of integrated 
luminosity for benchmark parameter choices relevant to cosmology. 
Our analysis directly maps onto the decoupling limits of the Next-to-Minimal 
Supersymmetric Standard Model (NMSSM) and more generically onto extensions of the Standard Model Higgs sector in 
which a heavy scalar produced through gluon fusion decays to a pair of Standard Model-like Higgs bosons.
\end{abstract}

\maketitle

\section{Introduction.}
\label{sec:intro}

Both ATLAS and CMS observe a Standard Model-like Higgs boson with $\sim$125 GeV mass.
While ongoing analyses show that the properties  of the newly discovered particle
are close to those expected for the Standard Model (SM) Higgs boson $h$, the full structure of the 
scalar sector responsible for electroweak symmetry-breaking remains to be determined.
It is particularly interesting to ascertain whether the scalar sector consists of only one SU(2)$_L$ doublet ($H$) or
has a richer structure containing additional states. Addressing this question is an important task for future studies at the Large Hadron Collider.

An interesting avenue for the observation of additional scalar states $X$ occurs in Higgs portal scenarios that contain operators of the form
$X H^\dag H$ and $X^2 H^\dag H$. For $m_X> 2 m_h$, these operators enable the process $ p p \to X^0 \to h h$, where $X^0$ is the neutral component of 
$X$, if $X$ is not inert with respect to the Standard Model. 
Signatures of such resonant di-Higgs production are multiparticle final states comprised of the conventional Higgs boson decay products. Di-Higgs 
production also occurs purely within the SM, though it cannot receive any enhancement due to an intermediate resonance (for studies of Higgs self-coupling 
probes with di-Higgs production at the LHC, see \cite{Djouadi:1999rca,Baur:2002qd,Baur:2003gpa,Baur:2003gp,Dolan:2012rv,Papaefstathiou:2012qe,Goertz:2013kp,Barr:2013tda}).

Higgs portal scenarios are strongly motivated by cosmology. 
In the presence of a discrete $Z_2$ symmetry, $X^0$ may be a dark matter candidate. In this case, 
the cubic operator is forbidden, the vacuum expectation value (vev) of $X^0$ vanishes, and resonant di-Higgs production cannot occur. 
In the absence of a $Z_2$ symmetry, however, both the cubic operator and a non-vanishing $X^0$ vev can exist. Under these conditions, 
the presence of the $X^0$ may facilitate a strong first order electroweak phase transition (EWPT) as required by electroweak baryogenesis (EWBG) 
(for a recent review, see \cite{Morrissey:2012db}). In this case, one would encounter a pair of neutral mass eigenstates $h_{1,2}$ formed from mixtures of 
the two neutral scalar fields, and for $m_{h_2}\geq 2 m_{h_1}$ resonant di-Higgs production could proceed (see also \cite{Dolan:2012ac}).

In what follows, we investigate the prospects for observing such Higgs portal-mediated resonant di-Higgs production in the context of the simplest extension 
of the SM scalar sector involving one real gauge singlet, $S$. 
This \lq\lq xSM" scenario can give rise to a strong first 
order electroweak phase transition as needed for electroweak baryogenesis in regions of parameter space that would also enable resonant 
di-Higgs production \cite{Profumo:2007wc,Espinosa:2011ax}). 
Study of the xSM also allows for a relatively general analysis of Higgs portal mediated resonant di-Higgs production. 
In particular the present analysis can be  mapped directly onto the ``decoupling limit" $m_A \gg v$ of the Next-to-Minimal 
Supersymmetric Standard Model (NMSSM) \cite{Ellwanger:2009dp} as well other scenarios that include additional degrees of freedom 
not directly relevant to di-Higgs production. 

In this study, we concentrate on the $b{\bar b}\tau^+\tau^-$ final state, motivated in part by the analogous work on SM-only non-resonant 
di-Higgs production as well as by the considerations discussed in section IV\footnote{We thank B. Brau for suggesting the study of this final state to us.}. 
We find that with an appropriate strategy for background reduction, discovery of $h_2$ at the LHC may be feasible 
with $\sim$ 50 - 100 fb$^{-1}$. Other final states resulting from combinations of Higgs decay products may also provide promising probes 
of the Higgs portal through resonant di-Higgs production, and we defer an analysis of these possibilities 
to future work\footnote{As this paper was being prepared for submission, an investigation of these other states 
appeared in Ref.~\cite{Liu:2013woa}. The results of the latter analysis differ considerably from ours, as we discuss below.}. 

The discussion of our analysis leading to this conclusion is organized as follows. In Section \ref{sec:bsm} we 
review the theoretical framework and motivation for the xSM. Section \ref{sec:constraints} gives the present LHC constraints and discusses 
other phenomenological considerations. In Section \ref{sec:lhc} we discuss the details and present the results of our LHC simulations and analysis, while 
in Section \ref{sec:discuss} we discuss their implications.

\section{Singlet Scalars Beyond the SM}
\label{sec:bsm}

Singlet scalar extensions of the SM are both strongly motivated and widely studied \cite{O'Connell:2006wi}. In the present instance, 
we rely on the simplest version as a paradigm for Higgs portal interactions and the prospects for novel collider signatures. 
At the same time, singlet extensions of the scalar sector are interesting in their own right.
From a model-building perspective, singlet scalars arise 
in various SM extensions, such as those containing one or more additional U(1) groups that occur in string constructions or variants 
on the NMSSM. Cosmology provides additional motivation. As noted above, 
the presence of the singlet scalar can enable a strongly first order 
EWPT as needed for electroweak baryogenesis, while imposing a $Z_2$ symmetry on the potential allows the singlet 
scalar to be a viable dark matter candidate (for early references, see, {\em e.g.} Refs.~\cite{McDonald:1993ex,Burgess:2000yq}).
In principle, one may achieve both a viable dark matter candidate 
and a strongly first order EWPT for a complex scalar singlet extension in the presence of a spontaneously- and softly-broken 
global U(1) \cite{Barger:2008jx,Gonderinger:2012rd}. 

In what follows, we concentrate on the real singlet, though many of the features discussed below will apply to the real component of the 
complex singlet case as well. The corresponding scalar potential for the SM Higgs doublet $H$ and a real 
singlet scalar field $S$ is


\bea
\label{ScalarPotential1}
V(H, S) = -\mu^2 \left| H \right|^2 + \lambda \left| H \right|^4 + \frac{b_2}{2} S^2 + 
\frac{b_4}{4} S^4 \nonumber \\
+ \frac{a_2}{2} S^2 \left| H \right|^2 + \frac{a_1}{2} S \left| H \right|^2
+ \frac{b_3}{3} S^3 -  \frac{a_1\,v^2}{4} S
\eea

We note that the scalar potential of a general NMSSM in the decoupling regime $m_A \gg v$ ($m_A$ is the mass of the neutral pseudocsalar)
is precisely of the form (\ref{ScalarPotential1}) \cite{Ellwanger:2009dp}, so our analysis for the xSM could be directly mapped 
into that interesting scenario  (recent global fits of LHC data in the context 
of supersymmetric models tend to favor this regime \cite{Espinosa:2012in,Barbieri:2013hxa,Belanger:2013xza}). 
Studies of resonant di-Higgs production, though in a different context from the 
present one, have also been carried out \cite{Ellwanger:2013ova,Cao:2013si,Kang:2013rj}.

Following \cite{Barger:2007im}, we have 
incorporated the last term in (\ref{ScalarPotential1}) in order to cancel the 
singlet tadpole generated once the EW symmetry is spontaneously broken, with
\be
\label{HiggsVev}
H = \frac{1}{\sqrt{2}} \left(\begin{array}{c}
0 \\
h + v
\end{array}\right)
\ee
in the unitary gauge and with $v = 246$ GeV. Denoting the neutral component of $H$ by $H^0$, the minimization conditions 
$\partial V/\partial H^0=0$ and $\partial V/\partial S=0$ with lead to
\bea
\label{eq:mincond}
 H^0\bigl[-2\mu^2+4\lambda (H^0)^2 + a_2 S^2&+& a_1S\bigr] = 0\\
 \nonumber
 S\left[b_2+ b_3 S + b_4 S^2+ a_2(H^0)^2\right] &=& \frac{a_1}{2}\left[v^2/2-(H^0)^2\right]\ \ .
 \eea
For positive $b_{2-4}$ and $a_2$, $H$ as given in (\ref{HiggsVev}),  and  $\lambda v^2=\mu^2$ as in the Standard Model, the scalar singlet
does not develop a zero-temperature vev\footnote{Note that in \cite{Profumo:2007wc}, the finite-temperature analysis was performed for 
a potential not having the linear term in $S$; mapping from one case to the other amounts to performing a linear shift in the field 
$S$ at zero temperature.}.  
The resulting mass term in the potential is
\be
V_\mathrm{mass} = \frac{1}{2}\left(
\begin{array}{cc}
h & S 
\end{array}\right)
\left(\begin{array}{cc}
\lambda v^2 & a_1 v/2 \\
a_1 v/2 & b_2+a_2 v^2/2
\end{array}\right)
\left(\begin{array}{c}
h\\ S
\end{array}\right)\ .
\ee
The states $h$ and $S$ will mix after EWSB if $a_1 \neq 0$, with mixing 
angle denoted  by $\theta$. The mass eigenstates 
 $h_{1,2}$  can then be expressed in terms of $h$ and $S$ as 

\be
\label{ScalarPotentialDefinitions4}
\left(\begin{array}{c}
h_1\\ h_2
\end{array}\right) = 
\left(\begin{array}{cc}
c_\theta & s_\theta\\
-s_\theta & c_\theta
\end{array}\right)\,
\left(\begin{array}{c}
h\\ S
\end{array}\right)\ \ \ ,
\ee
where $c_\theta\equiv\cos\theta$ and $s_\theta\equiv\sin\theta$ with
\be
\label{eq:theta}
\tan\theta = \frac{x}{1+\sqrt{1+x^2}}
\ee
and 
\be
x=\frac{a_1v}{(\lambda-a_2/2)v^2-b_2}\ \ \ .
\ee
The corresponding masses are
\bea
\label{eq:masses}
m_{\pm}^2 &=& \frac{1}{2}\left[(\lambda+a_2/2)v^2+b_2\right] \\
\nonumber
&&\pm \frac{1}{2}{\bigl\vert} (\lambda-a_2/2)v^2-b_2 {\bigr\vert} \sqrt{1+x^2}
\eea
with $m_2 = m_{+}$ and $m_1= m_{-}$.

The scalar potential (\ref{ScalarPotential1}) may then be written in terms of the following seven independent parameters:
the two scalar masses $m_{1,2}$; the mixing angle  $\theta$;  $v$, $a_2$, $b_3$ and $b_4$. Henceforth, we assume that 
$h_1$ is the Higgs-like state currently being observed at the LHC, with $m_1 = 125$ GeV, and $h_2$ is a heavier scalar 
state with $m_2 > 2 m_1$. The quartic coupling $b_4 > 0$ is needed to assure stability of the potential along 
the $S$ direction. The value of the effective trilinear $h_2 h_1 h_1$ coupling
\bea
\nonumber
\label{eq:lambda211}
\lambda_{211} = b_3 s_\theta^2 c_\theta + a_2v s_\theta(c_\theta^2-s_\theta^2/2)\\\
+\frac{a_1}{4} c_\theta(c^2_\theta-2s^2_\theta)-3\lambda v c^2_\theta s_\theta
\eea
is clearly of vital importance to our analysis. Note that $\lambda$ and $a_1$ are implicitly  functions of $m_{1,2}$, $\theta$, $v$ and $a_2$ {\em via} 
Eqs.~(\ref{eq:theta}-\ref{eq:masses}).

\begin{figure}[tbp!]
\begin{center}
\includegraphics[width=0.45\textwidth]{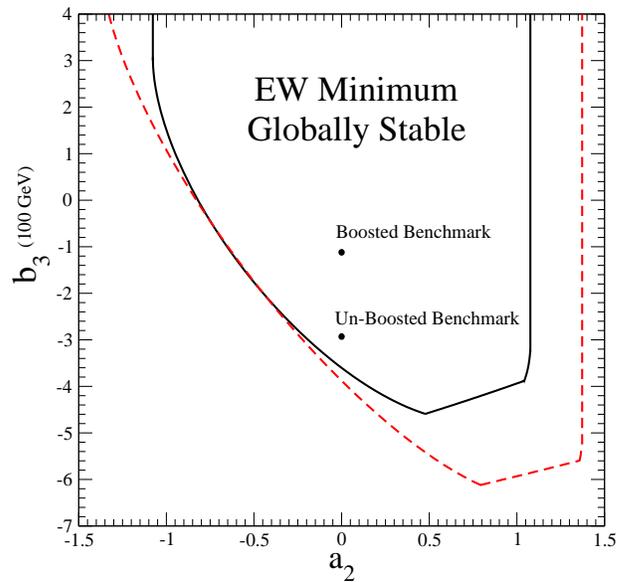}
\caption{\small Absolute stability region for the EW vacuum in the $(a_2, b_3)$ plane for $b_4 = 1$, $c_\theta = 0.812404$ and $m_2 = 270$ GeV (solid-black), 
$m_2 = 370$ GeV (dashed-red). The black dots correspond to the benchmark scenarios used in the analysis (see section IV).}
\label{fig:vacuum}
\end{center}
\end{figure}

Considerations of the vacuum structure of the potential introduce constraints on the independent parameters of the potential. Tree-level stability 
for large values of the fields $h$ and $S$ is ensured for positive $\lambda$, $b_4$ and $a_2$. However, allowing 
$a_2 < 0$ 
can enable a strong first order electroweak phase transition \cite{Profumo:2007wc}. 
In this case, requiring
$4\lambda b_4> a_2^2$ maintains stability of the potential\footnote{A strong first order EWPT can also occur for $a_2\geq 0$ for non-vanishing $a_1$.}.
This criterion becomes dependent on the cut-off of the low-energy effective theory after one 
takes into account the renormalization group evolution of the parameters, a consideration that 
we do not implement here (see, {\em e.g.}, \cite{Gonderinger:2012rd} and references therein). Note that for $a_2<0$ and/or $H$ different 
from (\ref{HiggsVev}), one may encounter additional solutions to (\ref{eq:mincond}) for which the $S$ vev does not vanish.  We require that if such 
additional minima exist, the $\langle S\rangle = 0$ extremum is a the global minimum. As illustrated in Fig.~\ref{fig:vacuum}, doing so leads to the
constraints in the $(a_2, b_3)$ plane for given values of $m_{1,2}$, $\theta$, and $b_4$. 
From Eq.~(\ref{eq:lambda211}) and the global stability region of Fig.~\ref{fig:vacuum}, we then observe that for each value of $a_2$ there exists a 
minimum value of $\lambda_{211}$ consistent with the vacuum structure requirements.

\section{Current  Constraints}
\label{sec:constraints}

For the process gluon-fusion mediated process $p p \rightarrow h_2 \rightarrow h_1 h_1$ of interest here, the magnitude of the cross section depends 
critically on the mixing
angle $\theta$ through both the $h_2$ coupling to SM quarks and the triscalar coupling $\lambda_{211}$. The mixing angle is 
constrained by the current LHC results for properties of the SM Higgs boson. On the one hand, 
the cross section for $p p \rightarrow h_1$ is reduced compared to 
the one for a 125 GeV SM Higgs by a factor $c_\theta^2$ due to the singlet-doublet mixing. 
On the other, although the coupling of $h_1$ to its decay products is also universally suppressed by 
$c_\theta$, its decay branching ratios are the same as for a SM Higgs since no new decay channels are open. 
Consequently, the observation of the SM-like Higgs at the LHC can  be used to set a lower bound on $c_\theta^2$ due to the associated signal 
suppression in SM Higgs decay channels.
Recent global analyses of LHC Higgs measurements then yield $c_\theta^2> 0.66$ at 0.95\% C.L. \cite{Winslow13,ATLASCoup}. 
From the analysis in \cite{Profumo:2007wc} we observe that for mixing angles in this range and  $m_2 > 2 m_1$, 
the xSM can lead to a strong first order electroweak phase transition.

Global fits to electroweak precision data also imply constraints on the mixing angle and $m_2$. Although a reanalysis of these constraints 
goes beyond the scope of the present investigation, previous studies indicate that significant singlet-doublet mixing is 
disfavored for heavier $h_2$ \cite{Profumo:2007wc}.

Another important constraint comes from ATLAS \cite{ATLASHeavyWW,ATLASHeavyZZ} and CMS \cite{CMSHeavy} direct searches for heavy scalars decaying to $W\,W$ and $Z\,Z$. As the resulting constraints are dependent on the heavy scalar mass, we note that in the next section we will choose as benchmark scenarios for our analysis $m_2 = 270$ GeV and $m_2 = 370$ GeV.
ATLAS searches in the $W\,W$ channel exclude $h_2$ at $95\, \%$ C.L. for $(\sigma \times \mathrm{Br})/(\sigma \times \mathrm{Br})_{\mathrm{SM}} \gtrsim 0.7$  
for $m_{2} \sim 270$ GeV and $(\sigma \times \mathrm{Br})/(\sigma \times \mathrm{Br})_{\mathrm{SM}} \gtrsim 0.4$ for $m_{2} \sim 370$ GeV, 
while $Z\,Z$ searches exclude $h_2$ at $95\, \%$ C.L. for $(\sigma \times \mathrm{Br})/(\sigma \times \mathrm{Br})_{\mathrm{SM}} \gtrsim 0.25$  
for $m_{h_2} \sim 270$ GeV and $(\sigma \times \mathrm{Br})/(\sigma \times \mathrm{Br})_{\mathrm{SM}} \gtrsim 0.3$ for $m_{h_2} \sim 370$ GeV. 
The bounds extracted from CMS searches are found to be similar. The production cross section for $h_2$ in the present case is given by 
$s_\theta^2 \, \sigma_{\mathrm{SM}}$, and thus for $s_\theta^2 \leq 0.34$ the constraints from $W\,W$ searches are satisfied, while 
a mild reduction in the branching fraction $h_2 \rightarrow Z\, Z$ compared to the SM, 
due to the $h_2 \rightarrow h_1\,h_1$ decay channel being available, suffices to satisfy also the constraints from $Z\, Z$ searches.   

\section{Resonant di-Higgs Production at the LHC}
\label{sec:lhc}

We now consider in detail resonant di-Higgs production at the LHC for $\sqrt{s}=14$ TeV. We focus  on 
the gluon fusion production mechanism that
is by far the dominant one for $m_2$ in the mass range of interest for the EWPT\footnote{We defer a study of 
associated production, weak boson fusion, and $t{\bar t}h_2$ production to future work.}. The production mechanism is analogous to Higgs pair production in the SM 
via the trilinear Higgs self-coupling \cite{Dolan:2012rv}, except that (a) the $s$-channel $ gg\to h_2 \to h_1h_1$ amplitude may be resonant in the 
present case (see also \cite{Dolan:2012ac}); and (b) the $ggh_2$ interaction will be reduced in strength by $c_\theta$. 

Before discussing our rationale for focusing on the $b{\bar b}\tau^+\tau^-$ final state, it is useful to compare the expected magnitudes of 
the resonant and non-resonant di-Higgs production cross sections for the ranges of masses and couplings we consider below.
The two most important contributions to the non-resonant cross section arise 
from the $gg\to h_1 h_1$ amplitude involving the top quark box graph and from the $gg\to h_1^\ast\to h_1 h_1$ process. The former will be reduced in strength from its 
SM value by $c^2_\theta$, while the latter will be reduced by 
$ c_\theta\times\lambda_{111}/\lambda_\mathrm{SM}$.  Taking $c^2_\theta=0.66$ and the SM di-Higgs production cross section 
from \cite{Dolan:2012rv} for $\lambda_{111}=0$ we obtain $\sigma(pp\to h_1 h_1)_\mathrm{non-res} \approx 26$ fb, which lies 
well below our typical values for the resonant cross section: $\mathcal{O}(1)\, \mathrm{pb}$ for $m_2 \lesssim 400$ GeV. Depending on the choices 
of the remaining independent parameters, the non-resonant $gg\to h_1^\ast\to h_1 h_1$ process may interfere constructively with the box contribution, 
leading to as much as a factor of two increase in the total non-resonant cross section. The resulting cross section nevertheless lies well below the typical 
resonant production cross sections for the range of $m_2$ that we study here, so we 
may safely disregard the non-resonant $h_1h_1$ contribution in our analysis.

For the signal, we consider the $b \bar{b} \tau^+ \tau^-$ final state since it has a sufficiently large branching ratio to yield a 
significant number of events with $\sim 100\,\, \mathrm{fb}^{-1}$ integrated luminosity yet does not contend with insurmountable backgrounds. 
For the 
final states with the largest branching ratio, $b \bar{b} b \bar{b}$ and $b{\bar b} W^+ W^-$, 
the substantial backgrounds ($\gsim 21$ pb and $\lsim 900$ pb cross sections, 
respectively, \cite{Dolan:2012rv}) are challenging at best and may be 
insurmountable\footnote{Recent analyses of generic resonant double SM-like Higgs production in the $b \bar{b} b \bar{b}$ suggest that it might be actually possible to 
efficiently suppress the large $b \bar{b} b \bar{b}$ QCD background using jet-substructure techniques \cite{Gouzevitch:2013qca}.}. In contrast, 
for the $b{\bar b}\tau^+\tau^-$ channel the potential $\lsim 900$ pb $b{\bar b} W^+W^-$ background gets reduced 
to $\lsim 20$ pb due to the small $W \rightarrow \ell \nu, \tau \nu$ branching fraction, as shown in studies of this channel in 
the context of SM di-Higgs production
Another potentially promising search channel is the $b{\bar b}\gamma\gamma$ final state. An earlier analysis of this channel in the 
context of the real triplet extension of the SM \cite{FileviezPerez:2008bj} indicates that discovery with $\sim 100$ fb$^{-1}$ of integrated 
luminosity would be possible  using these final state when the triplet scalar pair production cross section is of order one picobarn. 
As indicated above, we defer an investigation of this channel to future work. 

\begin{figure}[tbp!]
\begin{center}
\includegraphics[width=0.45\textwidth]{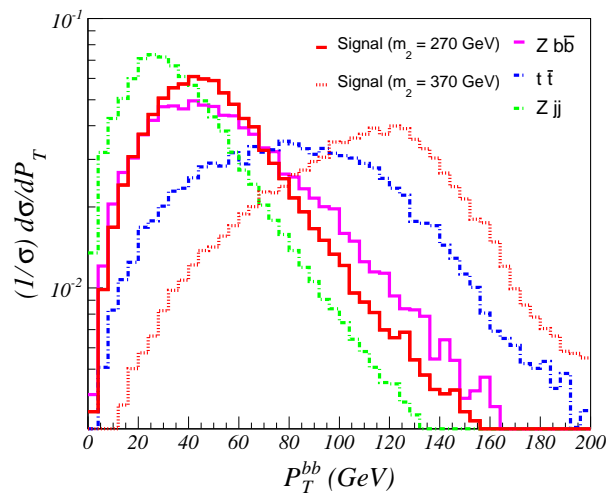}
\caption{\small Normalized $|{\vec P}_T|$ distribution for the $b \bar{b}$ system, for both signal and the dominant backgrounds.}
\label{fig:ptdist}
\end{center}
\end{figure} 

For the simulation of resonant di-Higgs production, we include both the $ gg\to h_2 \to h_1h_1$ and $ gg\to h_2+j \to h_1h_1+j$ processes in order to 
improve the reliability of the kinematic distributions of the $h_1$ bosons and their decay products, even though we do not explicitly make use of the presence of this 
additional hard jet in our analysis. For the  partonic gluon fusion process, 
we have implemented the xSM Lagrangian together with the scalar potential (\ref{ScalarPotential1}) in \textsf{FeynRules} \cite{Christensen:2008py,Degrande:2011ua}, 
including the $5$-dimensional gluon fusion effective operator $\mathcal{A}_{g} H G_{\mu\nu}^a G^{a\, \mu\nu}$
with the full LO form factor $\mathcal{A}_{g}$ that receives its leading 
contribution from the top quark triangle loop. Signal events are generated in \textsf{MadGraph/MadEvent 5} \
\cite{Alwall:2011uj} and subsequently interfaced to 
\textsf{Pythia} \cite{Sjostrand:2007gs} for parton showering, jet matching and hadronization using the CTEQ611 parton luminosities \cite{Pumplin:2002vw} set.
The events are finally interfaced to \textsf{PGS}, which uses an anti-$k_t$ jet reconstruction algorithm.
To set the overall normalization, we rescale our simulated K-factor
$K=\sigma(pp\to h_1h_1 X)_\mathrm{NLO}/\sigma(pp\to h_1h_1 X)_\mathrm{LO}$ by the value computed in Ref.~\cite{Dawson:1998py} and updated in 
Ref.~\cite{Baglio:2012np} that takes into account the  full set of NLO QCD corrections. 

We perform our study for two benchmark parameter space points: 
\begin{itemize}
\item[(a)] \textsl{Un-boosted Scenario}: $m_2 = 270$ GeV, $c_\theta = 0.812404$, $\lambda_{211}\, v = 325$ GeV (with $a_2 = 0$, $b_3 = -293$ GeV).
\item[(b)] \textsl{Boosted Scenario}: $m_2 = 370$ GeV, $c_\theta = 0.812404$, $\lambda_{211}\, v = 325$ GeV (with $a_2 = 0$, $b_3 = -112$ GeV).
\end{itemize}
For case (a), the di-Higgs pair is produced nearly at rest in the $h_2$ rest frame, so the $p_T$ distribution for each $h_1$ is peaked well below 150 GeV, 
as illustrated in Fig.~\ref{fig:ptdist}. There we show the $|{\vec p}_T|$ distribution of the $b{\bar b}$ pair produced by one of the decaying $h_1$ 
bosons along with the corresponding dominant backgrounds (see below).  For this regime, the results obtained from the effective theory above are expected to 
agree qualitatively very well with those using the full 1-loop matrix element \cite{Dolan:2012rv}. For case (b) the di-Higgs pair is 
boosted, with the $h_1$ $p_T$ distribution peaking near 130 GeV (see Fig.~~\ref{fig:ptdist}). In this regime, one approaches the limit of validity of the effective 
theory, so we do not consider heavier $m_2$. After taking int account NLO QCD corrections as discussed above, the corresponding inclusive di-Higgs production 
cross sections are 808 fb (420 fb) for the unboosted (boosted) scenarios.




\subsection{Analysis of $b \bar{b} \tau^+ \tau^-$ Final States}
\label{sec:analysis}

Maximizing the sensitivity to the $b{\bar b}\tau^+\tau^-$ produced from $h_2\rightarrow h_1 h_1$ decays entails reducing backgrounds 
generated by SM QCD and electroweak processes. A crucial step in this direction is the reconstruction of the invariant mass 
of the $b \bar{b}$ and $\tau^+ \tau^-$ pairs that should individually reproduce the $h_1$ peak. 
The MMC technique \cite{Elagin:2010aw} commonly used by the ATLAS and CMS collaborations \cite{Chatrchyan:2012vp,Aad:2012mea} to 
reconstruct the invariant mass of 
a $\tau^+ \tau^-$ system from a decaying resonance relies on maximum likelihood methods that are not possible to implement in the present analysis. Alternatively,
we use the collinear approximation \cite{Ellis:1987xu} to reconstruct the di-tau
invariant mass, which is used in experimental analyses of boosted resonances \cite{Aad:2012mea}. 
This procedure consists of assuming that the invisible neutrinos from 
the $\tau$ decays are emitted collinear with the visible products of the decay. It is then possible to obtain the absolute value of the 
missing momentum in each $\tau$ decay $p^{\mathrm{mis}}_{1,2}$ using the missing energy vector $\vec{E}^{\mathrm{miss}}_T$ 
in the event and the kinematics of the visible decay products:

\bea
\label{CollinearApp}
p^{\mathrm{mis}}_{1} = \frac{\mathrm{sin}(\phi^{\mathrm{vis}}_{2}) E^{\mathrm{miss}}_{Tx} - 
\mathrm{cos}(\phi^{\mathrm{vis}}_{2}) E^{\mathrm{miss}}_{Ty}}{\mathrm{sin}(\theta^{\mathrm{vis}}_{1}) \,
\mathrm{sin}(\phi^{\mathrm{vis}}_{2}-\phi^{\mathrm{vis}}_{1})} \\
p^{\mathrm{mis}}_{2} = \frac{\mathrm{cos}(\phi^{\mathrm{vis}}_{1}) E^{\mathrm{miss}}_{Ty} - 
\mathrm{sin}(\phi^{\mathrm{vis}}_{1}) E^{\mathrm{miss}}_{Tx}}{\mathrm{sin}(\theta^{\mathrm{vis}}_{2}) \,
\mathrm{sin}(\phi^{\mathrm{vis}}_{2}-\phi^{\mathrm{vis}}_{1})}
\eea
One then defines:
\be 
\label{CollinearApp2}
x_{1,2} = \frac{p^{\mathrm{vis}}_{1,2} }{p^{\mathrm{vis}}_{1,2} + p^{\mathrm{mis}}_{1,2}}
\ee 
where $p^{\mathrm{vis}}$ is the absolute 
value of the momentum of the visible products in each $\tau$ decay.
The invariant mass of the $\tau^+ \tau^-$ pair is then obtained as 
$m_{\tau\tau}^{\mathrm{coll}} = m_{\tau\tau}^{\mathrm{vis}}/\sqrt{x_1\,x_2}$, with $m_{\tau\tau}^{\mathrm{vis}}$
being the invariant mass of the visible decay products of the $\tau^+ \tau^-$ system.

The primary disadvantage of the collinear approximation (\ref{CollinearApp})-(\ref{CollinearApp2}) is that it is 
not well-defined when the two $\tau$'s from the 
decay of $h$ are emitted back-to-back in the transverse plane ($\left|\phi_1 - \phi_2 \right| \sim \pi$), which 
manifests itself in the divergence of $p^{\mathrm{mis}}_{1,2}$ as $\left|\phi_1 - \phi_2 \right| \rightarrow \pi$.
Moreover, in this configuration, 
the transverse momenta of the two  neutrinos
will tend to cancel each other, generically resulting in little missing energy ${E}_T^\mathrm{miss}$, which also renders the 
collinear approximation inefficient.

\begin{table}[t] 
\begin{center}
\begin{small}
\begin{tabular}{||c|c||} 
\hline
Description & Rationale    \\
\hline
$N_{b_{\mathrm{tag}}} = 2$ , $N_\ell = 2$ & signal selection \\
$p_T^\ell> 10$ GeV& lepton selection\\
$ p_T^b > 10$ GeV& $b$-jet selection\\
$\Delta R_{bb} > 0.5$, $|y_b| < 2.5$& $b$-jet selection\\
\hline
$\Delta R_{bb} > 2.1$ &  $Z jj$, $Zb{\bar b}$, $t{\bar t}$ reduction$^a$ \\ 
$P_{T,b_1} > 45 \,\mathrm{GeV}, \, P_{T,b_2} > 30  \,\mathrm{GeV}$&  $Z jj$, $Zb{\bar b}$, $t{\bar t}$ reduction$^b$ \\
$90  \,\mathrm{GeV} < m_{bb} < 140  \,\mathrm{GeV}$&  $h_1$ mass reconstruction$^c$ \\
Collinear $x_1,\,x_2$ Cuts & $m^{\mathrm{coll}}_{\tau\tau}$ reconstruction  \\
$\Delta R_{\ell\ell} > 2$ & $t{\bar t}$ reduction$^d$\\
$H_T^\mathrm{lept} < 120  \,\mathrm{GeV}$& $t{\bar t}$ reduction\\
$30  \,\mathrm{GeV}< m_{\ell\ell} < 75 \,\mathrm{GeV}$& $Z$-peak veto \\
$30  \,\mathrm{GeV}< m_{e \mu} < 100 \,\mathrm{GeV}$ &  \\
$100  \,\mathrm{GeV}< m^{\mathrm{coll}}_{\tau\tau} < 150 \,\mathrm{GeV}$& $h_1$ mass reconstruction \\
$E_T^\mathrm{miss} < 50  \,\mathrm{GeV}$ & $t{\bar t}$ reduction$^e$\\
$230 \,\mathrm{GeV}< m^{\mathrm{coll}}_{bb\tau\tau} < 300  \,\mathrm{GeV}$& $h_2$ mass reconstruction\\
 \hline
\end{tabular}
\end{small}
\end{center}
\caption[ . ]{Event selection criteria and ordered cut flow for background reduction in the $b{\bar b}\tau_\mathrm{lep}\tau_\mathrm{lep}$ channel. 
See: $^a$Fig.~\ref{fig:deltar}, $^b$Fig.~\ref{fig:ffPTb}, $^c$Fig.~\ref{fig:ffinvarmass},  $^d$Fig.~\ref{fig:deltar2}, and $^e$Fig.~\ref{fig:MET}.}
\label{tab:leplepcuts}
\end{table}

Imposing the collinear cut $0.1 < x_1, x_2 < 1$ eliminates events with a back-to-back configuration, so we use it 
when selecting events used for the reconstruction of the di-tau invariant mass, $m_{\tau\tau}^{\mathrm{coll}}$.
For the single Higgs gluon fusion process $p p \rightarrow h \rightarrow \tau^+ \tau^-$ 
the $\tau$ leptons are generically emitted nearly back-to-back since the Higgs is produced almost at rest 
in the transverse plane. The collinear approximation is more effective for single Higgs production in conjunction with a high-$p_T$ jet against which 
the di-tau pair recoils, thereby reducing the incidence of back-to-back $\tau$ pairs. For di-Higgs production, the $h_1$ decaying to the $b{\bar b}$ pair takes 
the place of the high $p_T$ jet, so we expect the use of the collinear approximation to be reasonably reliable in the case of our analysis (see also 
Ref.~\cite{Aad:2012mea}).

The most relevant backgrounds for the analysis of $b \bar{b} \tau^+ \tau^-$ final states are $Zb{\bar b}$, 
$Z+jets$ (with two jets mis-identified as $b$ quark objects) and $t {\bar t}$ production (the primary source of the large $b{\bar b} W^+ W^-$ background indicated above). 
As we do not consider in the present analysis the possibility of jets faking hadronically decaying $\tau$ leptons, 
we disregard certain possible (albeit less important) backgrounds such as $b{\bar b}Wj$ and $b{\bar b}jj$.
As with the signal, all background events are generated in \textsf{MadGraph/MadEvent 5} and subsequently interfaced to 
\textsf{Pythia} and \textsf{PGS}. The various background cross-sections are normalized to their respective NLO values 
via enhancement $K$ factors: $K \simeq 1.4$ for $Zb{\bar b}$ \cite{Campbell:2005zv} and $K \simeq 1.5$ for $t {\bar t}$ \cite{Mangano:1991jk,Bevilacqua:2010qb}  
(for $Z jj$, the NLO cross section is similar to the LO one for renormalization and factorization scales chosen as 
$\mu_R = \mu_F = M_Z$ \cite{Campbell:2003hd}). Following \cite{ATLAS_Tech}, our detector simulation is normalized to a 70\% b-tagging efficiency for 
b-quark jets with $|y| < 2.5$ together with a 60\% efficiency for identification of hadronic $\tau$'s.

\begin{table*}[tbp!]
\begin{tabular}{l|c|c c c| c | c | }

& $h_2 \rightarrow h_1 h_1$& & $t\bar t$ & & $Z\, b\bar b$& $Z\, jj$ \\
\hline
& $b{\bar b}\tau_\mathrm{lep}\tau_\mathrm{lep}$& $b{\bar b}\ell\ell$& $b{\bar b}\ell\tau_\mathrm{lep}$& $b{\bar b}
\tau_\mathrm{lep}\tau_\mathrm{lep}$& $b{\bar b}\ell\ell$ + $b{\bar b}
\tau_\mathrm{lep}\tau_\mathrm{lep}$ & $jj\ell\ell$ + $jj
\tau_\mathrm{lep}\tau_\mathrm{lep}$ \\
\hline
Event selection (see section V.B)& 7.47  & 11209 & 4005  & 289 & 8028 & 1144 \cr
$\Delta R_{bb} > 2.1$, $P_{T,b_1} > 45$ GeV, $P_{T,b_2} > 30$ GeV& 4.46  & 5585 & 2013 & 145 & 2471 & 153 \cr
$h_1$-mass: $90$ GeV $< m_{bb} < 140$ GeV& 3.12  & 1073 & 405 & 30 & 880 & 47 \cr
Collinear $x_1,\,x_2$ Cuts& 2.34 & 438 & 164 & 14.1  & 248 &  18  \cr
$\Delta R_{\ell\ell} > 2$, $H_T^\mathrm{lept} < 120$ GeV& 2.08 & 226 & 82 & 7.9 & 200 & 16.7 \cr
$30$ GeV $< m_{\ell\ell}\, (m_{e \mu}) < 75\, (100)$ GeV& 1.86  & 136 & 49 & 5.7 & 11.6 &  0.95  \cr
$h_1$-mass: $100$ GeV $< m^{\mathrm{coll}}_{\tau\tau} < 150$ GeV& 1.05 & 32.5 & 11.4 & 1.63 & 3.24  & 0.24 \cr
$E_T^\mathrm{miss} < 50$ GeV& 0.89 & 10.5 & 3.37 & 0.56 & 3.03 & 0.23 \cr
$h_2$-mass: $230$ GeV $< m^{\mathrm{coll}}_{bb\tau\tau} < 300$ GeV& 0.81 & 1.19 & 0.39 & 0.12 & 0.86 & 0.09 \cr
\hline
\end{tabular}
\caption{
Event selection and background reduction for the $b{\bar b}\tau_\mathrm{lep}\tau_\mathrm{lep}$ channel in the \textsl{un-boosted} benchmark scenario. 
We show the NLO cross section (in $\mathrm{fb}$) for the signal $h_2 \rightarrow h_1 h_1 \rightarrow b{\bar b}\tau_\mathrm{lep}\tau_\mathrm{lep}$ 
and the relevant backgrounds $t\bar t \rightarrow b{\bar b}\tau_\mathrm{lep}\tau_\mathrm{lep},\, b{\bar b}\ell\tau_\mathrm{lep},\,b{\bar b}\ell\ell$,
$Z\, b\bar b \rightarrow b{\bar b}\tau_\mathrm{lep}\tau_\mathrm{lep},\, \,b{\bar b}\ell\ell$ and 
$Z\, jj \rightarrow jj\tau_\mathrm{lep}\tau_\mathrm{lep},\, \,jj\ell\ell$ after successive cuts. 
A 70\% b-tagging efficiency is assumed, following \cite{ATLAS_Tech}, together with a jet fake 
rate of 2\% (slightly more conservative than that from \cite{ATLAS_Tech}).}
\label{tab:leplepcutflow1}
\end{table*}

\begin{table*}[tbp!]
\begin{tabular}{l|c|c c c| c | c | }

& $h_2 \rightarrow h_1 h_1$& & $t\bar t$ & & $Z\, b\bar b$& $Z\, jj$ \\
\hline
& $b{\bar b}\tau_\mathrm{lep}\tau_\mathrm{lep}$& $b{\bar b}\ell\ell$& $b{\bar b}\ell\tau_\mathrm{lep}$& $b{\bar b}
\tau_\mathrm{lep}\tau_\mathrm{lep}$& $b{\bar b}\ell\ell$ + $b{\bar b}
\tau_\mathrm{lep}\tau_\mathrm{lep}$ & $jj\ell\ell$ + $jj
\tau_\mathrm{lep}\tau_\mathrm{lep}$ \\
\hline
Event selection (see section V.B)& 4.24 & 11209 & 4005  & 289 & 8028 & 1144  \cr
$\Delta R_{bb} < 2.2$, $P_{T,b_1} > 50$ GeV, $P_{T,b_2} > 30$ GeV& 2.38  & 3356 & 1202 & 85 & 1166  & 35 \cr
$h_1$-mass: $90$ GeV $< m_{bb} < 140$ GeV& 1.89 & 1396 & 512 & 36  & 452 & 12 \cr
$|\vec{P}^{bb}_T| > 110$ GeV& 1.35 & 719 & 264 & 19 & 208 &  4.9 \cr
Collinear $x_1,\,x_2$ Cuts& 1.09 &  293 & 107 & 8.8 & 58 & 1.86 \cr
$\Delta R_{\ell\ell} < 2.3$, $H_T^\mathrm{lept} < 120$ GeV& 0.80 & 120 & 45  & 4.2 & 9  & 0.14 \cr
$30$ GeV $< m_{\ell\ell}\, (m_{e \mu}) < 75\, (100)$ GeV& 0.70 & 85 & 30 & 2.45 & 1.51 &  0.019 \cr
$h_1$-mass: $100$ GeV $< m^{\mathrm{coll}}_{\tau\tau} < 150$ GeV& 0.60 & 30 & 11 & 0.96 & 0.24 & 0.003\cr
$25$ GeV $ < E_T^\mathrm{miss} < 90$ GeV& 0.42 & 18& 6.2 & 0.60 & 0.18& 0.003 \cr
$h_2$-mass: $330$ GeV $< m^{\mathrm{coll}}_{bb\tau\tau} < 400$ GeV& 0.32 & 3.25 & 1.08  & 0.11 & 0.025 & $<$ 0.001\cr
\hline
\end{tabular}
\caption{
Event selection and background reduction for the $b{\bar b}\tau_\mathrm{lep}\tau_\mathrm{lep}$ channel in the \textsl{boosted} benchmark scenario (same assumptions as 
in Table \ref{tab:leplepcutflow1}).}
\label{tab:leplepcutflow2}
\end{table*}

It is useful to organize the analysis according to the different $\tau$-decay modes, following roughly the treatment in 
Ref. \cite{Aad:2012mea}. We, thus, consider $b{\bar b}$ plus (a) two leptonically decaying $\tau$s (\lq\lq $\tau_\mathrm{lep}\tau_\mathrm{lep}$"); 
(b) one leptonically decaying and one hadronically decaying  $\tau$ (\lq\lq $\tau_\mathrm{lep}\tau_\mathrm{had}$"); and (c) two hadronically 
decay $\tau$s (\lq\lq $\tau_\mathrm{had}\tau_\mathrm{had}$").  After $\tau$ identification and b-tagging, the NLO cross sections for the unboosted (boosted) case are:
(a) $\tau_\mathrm{lep}\tau_\mathrm{lep}$: $10.58 \,(5.75)\,\mathrm{fb}$; (b) $\tau_\mathrm{lep}\tau_\mathrm{had}$: $23.39 \, (12.71)\,\mathrm{fb}$; and 
(c) $\tau_\mathrm{had}\tau_\mathrm{had}$: $12.90 \,(7.01)\, \mathrm{fb}$ for a total cross section of $46.85 \,(25.48)\,\mathrm{fb}$.


\subsection{Leptonic ($\tau_\mathrm{lep}\tau_\mathrm{lep}$) final states.}
\label{subsec:sigback}

When the two $\tau$-leptons in the final state decay leptonically ($\tau_\mathrm{lep}\tau_\mathrm{lep}$), 
the relevant backgrounds are 
$t\bar t \rightarrow b{\bar b}\ell\ell\nu{\bar\nu}$, 
$t\bar t \rightarrow b{\bar b}\ell\tau_\mathrm{lep}\nu{\bar\nu}$,
$t\bar t \rightarrow b{\bar b}\tau_\mathrm{lep}\tau_\mathrm{lep}\nu{\bar\nu}$,
$Z\, b\bar b \rightarrow b{\bar b}\ell\ell$, 
$Z\, b\bar b \rightarrow b{\bar b}\tau_\mathrm{lep}\tau_\mathrm{lep}$,  
$Z\, jj \rightarrow jj\ell\ell$ and
$Z\, jj \rightarrow jj\tau_\mathrm{lep}\tau_\mathrm{lep}$.
A summary of our selection and background reduction cuts for the unboosted case appears in Table \ref{tab:leplepcuts}. 
For the boosted pair case as well as for the $\tau_\mathrm{lep}\tau_\mathrm{had}$ and $\tau_\mathrm{had}\tau_\mathrm{had}$ final states, 
we will subsequently discuss modifications of this basic set of cuts implemented in our analysis. 

\begin{figure}[tbp!]
\begin{center}
\includegraphics[width=0.45\textwidth]{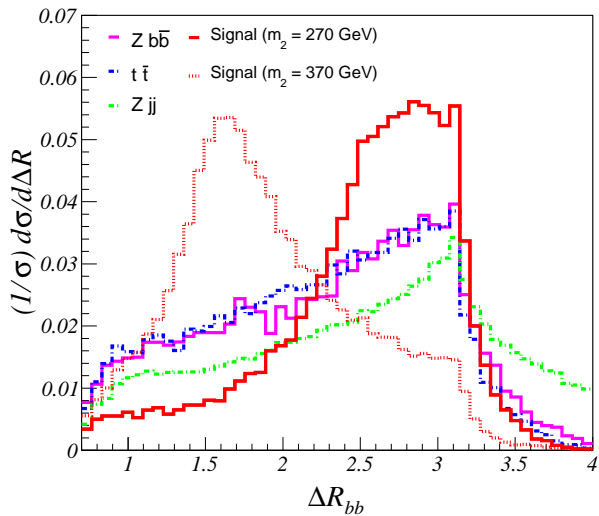}
\caption{\small Normalized $\Delta R_{bb}$ distribution after event 
selection (before cuts) for signal and background (\lq\lq $\tau_\mathrm{lep}\tau_\mathrm{lep}$").}
\label{fig:deltar}
\end{center}
\end{figure}

\begin{figure}[tbp!]
\begin{center}
\includegraphics[width=0.45\textwidth]{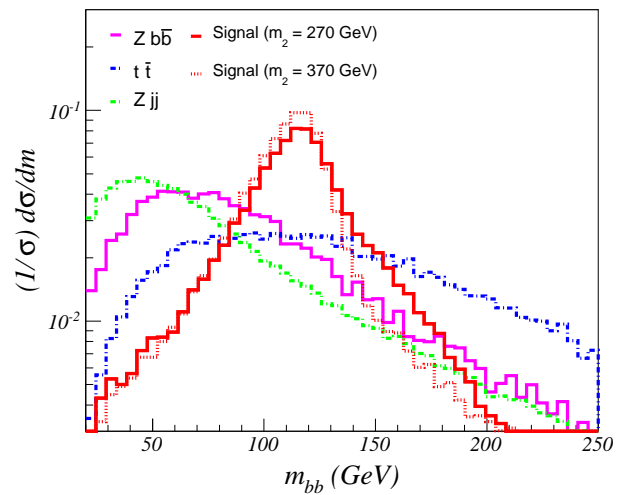}
\caption{\small Normalized $m_{bb}$ distribution after event 
selection (before cuts) for signal and background (\lq\lq $\tau_\mathrm{lep}\tau_\mathrm{lep}$").}
\label{fig:ffinvarmass}
\end{center}
\end{figure}

\begin{figure}[tbp!]
\begin{center}
\includegraphics[width=0.45\textwidth]{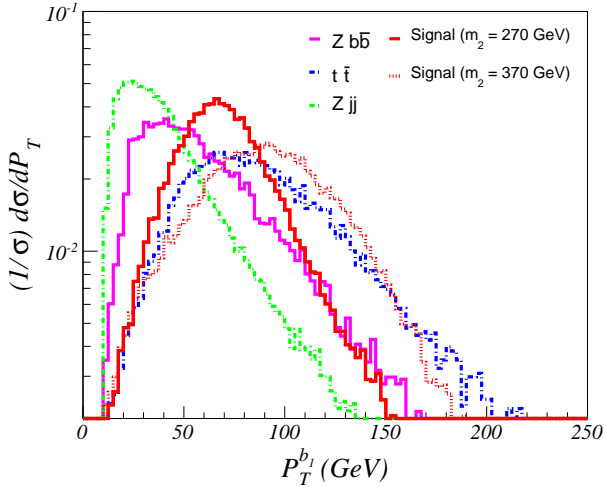}
\caption{\small Normalized $P_{T,b_1}$ distribution after event 
selection (before cuts) for signal and background (\lq\lq $\tau_\mathrm{lep}\tau_\mathrm{lep}$").}
\label{fig:ffPTb}
\end{center}
\end{figure}

\begin{figure}[tbp!]
\begin{center}
\includegraphics[width=0.45\textwidth]{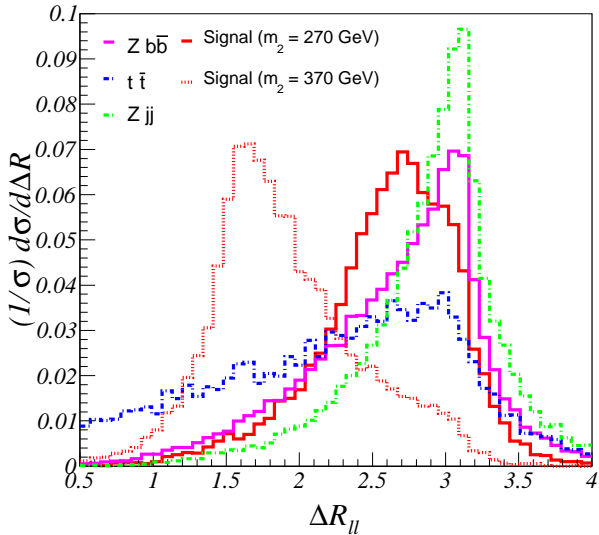}
\caption{\small Normalized $\Delta R_{\ell\ell}$ distribution after event 
selection (before cuts) for signal and background (\lq\lq $\tau_\mathrm{lep}\tau_\mathrm{lep}$").}
\label{fig:deltar2}
\end{center}
\end{figure}

For the analysis of the $\tau_\mathrm{lep}\tau_\mathrm{lep}$ channel we select events containing exactly 
two $b$-tagged jets ($N_{b_{\mathrm{tag}}} = 2$) and two isolated leptons ($N_\ell = 2$).
The cuts in $\Delta R_{bb}$, the $p_T$ of the two $b$-tagged jets and the invariant mass reconstructions for $b{\bar b}$, $\tau\tau$ 
and $b{\bar b}\tau\tau$  significantly reduce all backgrounds (see Figs. \ref{fig:deltar}, \ref{fig:ffinvarmass}, \ref{fig:ffPTb}). In addition,  
the $Z$ backgrounds can be further suppressed by imposing cuts on the dilepton 
invariant mass, while $t{\bar t}$ is suppressed with a combination of cuts on $E_T^\mathrm{miss}$, the $\Delta R$ of the 
reconstructed di-tau pair (see Fig. \ref{fig:deltar2}), 
and the scalar sum of leptonic transverse momentum, $H_T^\mathrm{lept}$. We include all possible combinations of opposite sign leptons in our simulated 
samples ($ee$, $e\mu$ and $\mu\mu$). Further reduction of the $Z$ backgrounds could be achieved by considering only $e\mu$ pairs as in 
Ref.~\cite{Aad:2012mea}. Doing so in the present case, however, leads to a loss of signal without significantly improving the final $S/\sqrt{S+B}$.  

For the boosted benchmark scenario, the $P_T$ of each $h_1$ will in general be substantially higher (see Fig.~\ref{fig:ptdist}), and the 
$h_1$ decay products will tend to be more collimated. We accordingly modify our cuts   
by imposing an upper bound on both $\Delta R_{bb}$ and $\Delta R_{\ell\ell}$ together with an increase on the $P_{T,b_1}$ threshold,
as suggested by Figs.~\ref{fig:deltar}, \ref{fig:ffPTb} and \ref{fig:deltar2}. 
While the $t{\bar t}$ distributions for $\Delta R_{bb}$ and $\Delta R_{\ell\ell}$ are relatively flat, those for the signal shift dramatically 
from the large to small $\Delta R$ range when going from the unboosted to the boosted regime (the $Zb{\bar b}$ and $Zjj$ backgrounds are reduced 
with separate cuts). In addition, we find further improvement in the $Zjj$ and $Z b {\bar b}$ background reduction
by requiring a relatively large $|\vec{P}^{bb}_T|$ as is apparent from Fig.~\ref{fig:ptdist}.
The corresponding impact of the cut-flow on signal and background cross sections are given in Tables \ref{tab:leplepcutflow1} and \ref{tab:leplepcutflow2}
for the unboosted and boosted scenarios, respectively.

\begin{figure}[tbp!]
\begin{center}
\includegraphics[width=0.45\textwidth]{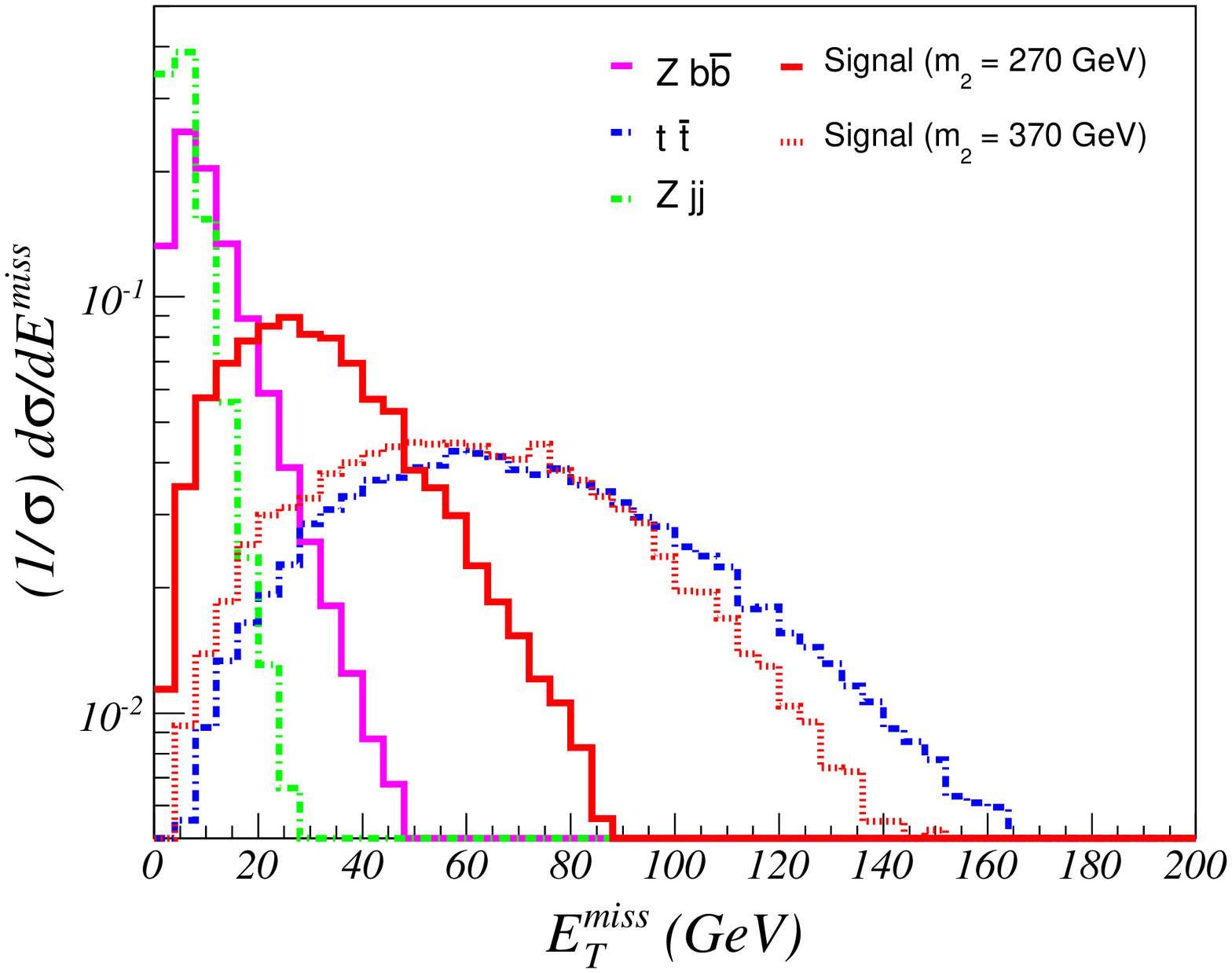}
\caption{\small Normalized $E_T^\mathrm{miss}$ distribution after event 
selection (before cuts) for signal and background (\lq\lq $\tau_\mathrm{lep}\tau_\mathrm{lep}$")}
\label{fig:MET}
\end{center}
\end{figure}

In light of the results from Tables \ref{tab:leplepcutflow1} and \ref{tab:leplepcutflow2}, for both $\tau$-leptons decaying leptonically a 
$S/\sqrt{S+B} \sim 5$ for the unboosted benchmark scenario can be achieved 
with $\sim 130 - 140 \, \mathrm{fb}^{-1}$, while the boosted benchmark scenario requires $\gtrsim 1000 \, \mathrm{fb}^{-1}$.
The inability to efficiently reduce the $t {\bar t}$ background in the latter case 
is related to the greater amount of $E_T^\mathrm{miss}$ in the signal events (coming from the decay of the more boosted $\tau$-leptons) 
for the boosted scenario, which then renders the cut on $E_T^\mathrm{miss}$ relatively inefficient in suppressing the $t {\bar t}$ background, 
in contrast to the situation in the un-boosted scenario (see Fig.~\ref{fig:MET}). 

\subsection{Semileptonic ($\tau_\mathrm{lep}\tau_\mathrm{had}$) final states.}
\label{subsec:sigback2}

\begin{table*}[tbp!]
\begin{tabular}{l|c|c c c| c | c | }

& $h_2 \rightarrow h_1 h_1$& & $t\bar t$ & & $Z\, b\bar b$& $Z\, jj$  \\
\hline
& $b{\bar b}\tau_\mathrm{lep}\tau_\mathrm{had}$& $b{\bar b}\ell\tau_\mathrm{had}$& & $b{\bar b}
\tau_\mathrm{lep}\tau_\mathrm{had}$& $b{\bar b} \tau_\mathrm{lep}\tau_\mathrm{had}$ & $jj \tau_\mathrm{lep}\tau_\mathrm{had}$  \\
\hline
Event selection (see section V.C)& 19.17 & 5249 &  & 762 & 601 & 98 \cr
$\Delta R_{bb} > 2.1$, $P_{T,b_1} > 45$ GeV, $P_{T,b_2} > 30$ GeV& 11.45  & 2639  & & 384  & 188 & 10.8 \cr
$h_1$-mass: $90$ GeV $< m_{bb} < 140$ GeV& 8.00  & 531 & & 80  & 69 & 3.68 \cr
Collinear $x_1,\,x_2$ Cuts& 4.81 & 209 & & 36.4  & 41.6 & 2.41  \cr
$\Delta R_{\ell\tau} > 2$& 4.10 & 129  &  & 23.1 & 26.5 &  2.03  \cr
$m^{\ell}_T < 30$ GeV& 3.44  & 30.9  &  & 11.1  & 24.4  &  1.90   \cr
$h_1$-mass: $110$ GeV $< m^{\mathrm{coll}}_{\tau\tau} < 150$ GeV& 1.56 &4.97 &  & 2.05  & 4.92 & 0.38  \cr
$E_T^\mathrm{miss} < 50$ GeV& 1.37 & 3.31 &  & 0.87 & 4.29 & 0.36 \cr
$h_2$-mass: $ 230$ GeV $< m^{\mathrm{coll}}_{bb\tau\tau} < 300$ GeV& 1.29 & 0.39 &  & 0.17  & 1.21 & 0.13\cr
\hline
\end{tabular}
\caption{
Event selection and background reduction for the $b{\bar b}\tau_\mathrm{lep}\tau_\mathrm{had}$ channel in the un-boosted benchmark scenario. 
We show the NLO cross section (in $\mathrm{fb}$) for the signal $h_2 \rightarrow h_1 h_1 \rightarrow b{\bar b}\tau_\mathrm{lep}\tau_\mathrm{had}$ 
and the relevant backgrounds $t\bar t \rightarrow b{\bar b}\tau_\mathrm{lep}\tau_\mathrm{had},\, b{\bar b}\ell\tau_\mathrm{had}$,
$Z\, b\bar b \rightarrow b{\bar b}\tau_\mathrm{lep}\tau_\mathrm{had}$ and 
$Z\, jj \rightarrow jj\tau_\mathrm{lep}\tau_\mathrm{had}$ after successive cuts (same efficiency and face rate assumptions as in Table \ref{tab:leplepcutflow1}). }
\label{tab:lephadcuts}
\end{table*}

\begin{table*}[tbp!]
\begin{tabular}{l|c|c c c| c | c | }

& $h_2 \rightarrow h_1 h_1$& & $t\bar t$ & & $Z\, b\bar b$& $Z\, jj$  \\
\hline
& $b{\bar b}\tau_\mathrm{lep}\tau_\mathrm{had}$& $b{\bar b}\ell\tau_\mathrm{had}$& & $b{\bar b}
\tau_\mathrm{lep}\tau_\mathrm{had}$& $b{\bar b} \tau_\mathrm{lep}\tau_\mathrm{had}$ & $jj \tau_\mathrm{lep}\tau_\mathrm{had}$  \\
\hline
Event selection (see section V.C)& 10.73 & 5249 &  & 762 & 601 & 98 \cr
$\Delta R_{bb} < 2.2$, $P_{T,b_1} > 50$ GeV, $P_{T,b_2} > 30$ GeV& 6.02  & 1576  &  & 223 & 85 & 2.46 \cr
$h_1$-mass: $90$ GeV $< m_{bb} < 140$ GeV& 4.77  & 672 & & 94  & 31.5  & 0.84 \cr
$|\vec{P}^{bb}_T| > 110$ GeV& 3.42  & 345 &  & 49 & 13.9 & 0.33 \cr
Collinear $x_1,\,x_2$ Cuts& 2.31  & 136  & & 22.3  & 8.38 & 0.22  \cr
$\Delta R_{\ell\tau} < 2.3$& 1.71 & 68 &  & 11.1 & 4.31 &  0.055  \cr
$m^{\ell}_T < 30$ GeV& 1.46  & 18.4  &  &  5.64 & 4.02  &  0.051   \cr
$h_1$-mass: $110$ GeV $< m^{\mathrm{coll}}_{\tau\tau} < 150$ GeV& 1.05 & 4.2 &  & 1.26 & 0.30 & 0.003  \cr
$25$ GeV $ < E_T^\mathrm{miss} < 90$ GeV & 0.76 & 2.93  &  & 0.75 & 0.23 & 0.002 \cr
$h_2$-mass: $ 330$ GeV $< m^{\mathrm{coll}}_{bb\tau\tau} < 400$ GeV& 0.63 & 0.60 &  & 0.15  & 0.026 & $<$ 0.001 \cr
\hline
\end{tabular}
\caption{
Event selection and background reduction for the $b{\bar b}\tau_\mathrm{lep}\tau_\mathrm{had}$ channel in the boosted benchmark scenario (same 
efficiency and face rate assumptions as in Table \ref{tab:leplepcutflow1}).}
\label{tab:lephadcuts_2}
\end{table*}

\begin{figure}[tbp!]
\begin{center}
\includegraphics[width=0.45\textwidth]{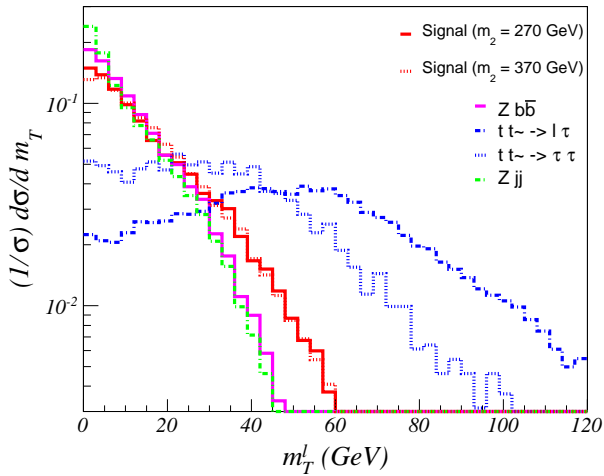}
\caption{\small Normalized $m^{\ell}_T $ distribution for signal and background (\lq\lq $\tau_\mathrm{lep}\tau_\mathrm{had}$").}
\label{fig:mT}
\end{center}
\end{figure}

For the $b{\bar b} \tau_\mathrm{lep}\tau_\mathrm{had}$ final state, we require exactly one isolated lepton and 
one hadronically decaying tau (\lq\lq $\tau_h$"), where the latter is identified using the PGS detector simulator. 
The event selection criteria for this channel are given by: $N_{b_{\mathrm{tag}}} = 2$, $N_\ell = 1$, $N_{\tau_h} = 1$ , $p_T^\ell, p_T^\tau> 10$ 
GeV, $|y_b| < 2.5$, $\Delta R_{bb} > 0.5$, $ p_T^b > 10$.
The main backgrounds arise from $t{\bar t}$ with $b{\bar b}\ell\tau_\mathrm{had}\nu{\bar\nu}$ and $b{\bar b}\tau_\mathrm{lep}\tau_\mathrm{had}\nu{\bar\nu}$
produced in the $t$-quark decays, and $Zb{\bar b}$, $Zjj$ with $Z\to \tau_\mathrm{lep}\tau_\mathrm{had}$.
The imposed cuts are similar to those applied to the $\tau_\mathrm{lep}\tau_\mathrm{lep}$ case, except for the di-lepton 
invariant mass cuts. Instead, to reduce backgrounds associated with $b{\bar b}WW$ events (largely dominated by $t{\bar t}$ production), 
we cut on the transverse mass of the lepton (see Fig. \ref{fig:mT})

\be
m^{\ell}_T=\sqrt{2 p_T^\ell E_T^\mathrm{miss} (1-\cos\phi_{\ell,\mathrm{miss}}) } < 30\ \mathrm{GeV}
\ee

with $\phi_{\ell,\mathrm{miss}}$ being the azimuthal angle between the direction of missing energy and the 
lepton transverse momentum. 

\begin{figure}[tbp!]
\begin{center}
\includegraphics[width=0.45\textwidth]{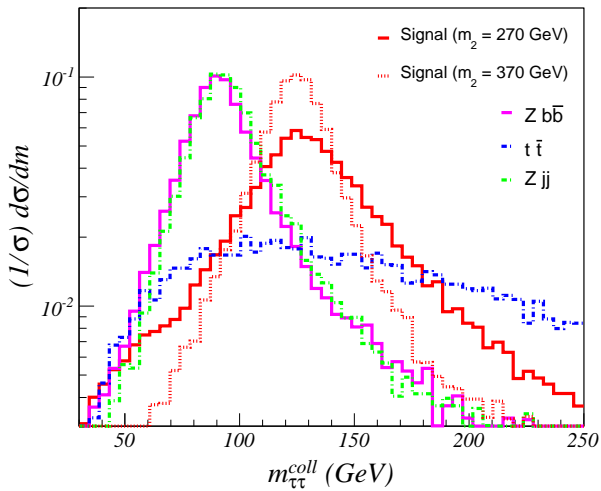}
\caption{\small Normalized $m^{\mathrm{coll}}_{\tau\tau}$ distribution for $\tau^+ \tau^-$ system in signal and 
background (\lq\lq $\tau_\mathrm{lep}\tau_\mathrm{had}$").}
\label{fig:mtautau}
\end{center}
\end{figure}

\begin{figure}[tbp!]
\begin{center}
\includegraphics[width=0.45\textwidth]{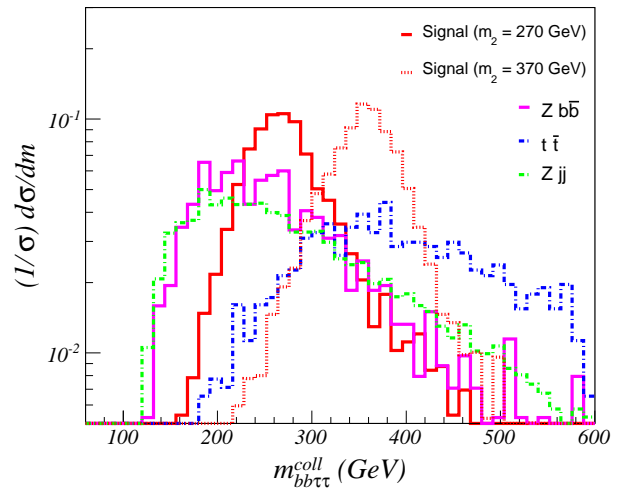}
\caption{\small Normalized $m^{\mathrm{coll}}_{bb\tau\tau}$ distribution for signal and 
background (\lq\lq $\tau_\mathrm{lep}\tau_\mathrm{had}$").}
\label{fig:mtautaubb}
\end{center}
\end{figure}

The corresponding impact of the cut-flow on signal and background cross sections are given in Tables \ref{tab:lephadcuts} and \ref{tab:lephadcuts_2}
for the unboosted and boosted scenarios. As for the $\tau_\mathrm{lep}\tau_\mathrm{lep}$ channel, the various cuts allow one to 
greatly suppress the backgrounds and increase the signal significance. For the $\tau_\mathrm{lep}\tau_\mathrm{had}$ channel, since it is not possible 
to impose a Z-peak veto through a cut in the invariant mass of the lepton pair, we increase the lower end of the $m^{\mathrm{coll}}_{\tau\tau}$ 
invariant mass signal window (from $100$ GeV to $110$ GeV) in order to suppress $Z b {\bar b}$ and $Z jj$ backgrounds. The distributions 
for $m^{\mathrm{coll}}_{\tau\tau}$ and $m^{\mathrm{coll}}_{bb\tau\tau}$ in this channel are shown in Figs. \ref{fig:mtautau} and \ref{fig:mtautaubb}.

From the results from Tables \ref{tab:lephadcuts} and \ref{tab:lephadcuts_2}, we find that for the semileptonic channel a 
$S/\sqrt{S+B} \sim 5$ for the unboosted benchmark scenario can be obtained 
with $\sim 50 \, \mathrm{fb}^{-1}$, while for the boosted benchmark scenario the required integrated luminosity is slightly higher, 
$\sim 90 \, \mathrm{fb}^{-1}$. This channel therefore appears to be promising both for the boosted and unboosted regimes.

\subsection{Hadronic ($\tau_\mathrm{had}\tau_\mathrm{had}$) final states.}
\label{subsec:sigback3}

The selection criteria for this channel are given by two hadronically-decaying $\tau$-leptons ($N_{\tau_h} = 2$), exactly zero leptons ( $N_{\ell} = 0$), and a 
similar set of kinematic requirements on the $\tau$ leptons and $b$-jets as in the other channels: $p_T^\tau> 10$ 
GeV, $|y_b| < 2.5$, $\Delta R_{bb} > 0.5$, $ p_T^b > 10$. As compared to the semileptonic and leptonic channels, the backgrounds for the purely 
hadronic channel are smaller. The cut flows for the unboosted and boosted scenarios are given in Tables \ref{tab:hadhadcuts} and \ref{tab:hadhadcuts_bis}, respectively. 

In light of the results from Tables \ref{tab:hadhadcuts} and \ref{tab:hadhadcuts_bis}, we obtain 
$S/\sqrt{S+B} \sim 5$ with $\sim 100 \, \mathrm{fb}^{-1}$ in the hadronic channel for both the unboosted and boosted benchmark scenarios. 
While this channel appears to be promising both for both scenarios, we caution that we have not considered other pure QCD backgrounds, 
such as multijet or $b {\bar b} jj$ production, where the jets fake a hadronically decaying $\tau$ lepton.  
The reason is the difficulty of reliably 
quantifying the jet fake rate for these events, which while being under $5\%$, 
depends strongly on the characteristics of the jet \cite{ATLAS_Tech}. While we do not expect this class of 
background contamination to be an impediment to signal observation in the $\tau_\mathrm{had}\tau_\mathrm{had}$ channel, we are less confident in 
our quantitative statements here than for the other final states. 

\begin{table*}[tbp!]
\begin{tabular}{l|c| c | c | c | }

& $h_2 \rightarrow h_1 h_1$& $t\bar t$ & $Z\, b\bar b$& $Z\, jj$  \\
\hline
& $b{\bar b}\tau_\mathrm{had}\tau_\mathrm{had}$& $b{\bar b}
\tau_\mathrm{had}\tau_\mathrm{had}$& $b{\bar b} \tau_\mathrm{had}\tau_\mathrm{had}$ & $jj \tau_\mathrm{had}\tau_\mathrm{had}$  \\
\hline
Event selection (see section V.D)& 12.31 & 509 & 411 & 67 \cr
$\Delta R_{bb} > 2.1$, $P_{T,b_1} > 45$ GeV, $P_{T,b_2} > 30$ GeV& 7.35 & 256  & 128 & 7.39  \cr
$h_1$-mass: $90$ GeV $< m_{bb} < 140$ GeV& 5.14  & 53 & 47  & 2.52 \cr
Collinear $x_1,\,x_2$ Cuts& 2.57 & 22.8 & 24.5 & 1.42  \cr
$\Delta R_{\tau\tau} > 2$& 2.04  & 12.4 & 15.8  & 1.19 \cr
$h_1$-mass: $110$ GeV $< m^{\mathrm{coll}}_{\tau\tau} < 150$ GeV& 0.82 & 1.79 & 3.75 & 0.27  \cr
$E_T^\mathrm{miss} < 50$ GeV& 0.75 & 0.60  & 3.39 &  0.26 \cr
$h_2$-mass: $ 230$ GeV $< m^{\mathrm{coll}}_{bb\tau\tau} < 300$ GeV&  0.72 & 0.08 & 1.03  & 0.11 \cr
\hline
\end{tabular}
\caption{
Event selection and background reduction for the $b{\bar b}\tau_\mathrm{had}\tau_\mathrm{had}$ channel in the unboosted benchmark scenario. 
We show the NLO cross section (in $\mathrm{fb}$) for the signal $h_2 \rightarrow h_1 h_1 \rightarrow b{\bar b}\tau_\mathrm{had}\tau_\mathrm{had}$ 
and the relevant backgrounds $t\bar t \rightarrow b{\bar b}\tau_\mathrm{had}\tau_\mathrm{had}$,
$Z\, b\bar b \rightarrow b{\bar b}\tau_\mathrm{had}\tau_\mathrm{had}$ and 
$Z\, jj \rightarrow jj\tau_\mathrm{had}\tau_\mathrm{had}$ after successive cuts (same efficiency and face rate assumptions as in Table \ref{tab:leplepcutflow1}). 
}
\label{tab:hadhadcuts}
\end{table*}

\begin{table*}[tbp!]
\begin{tabular}{l|c| c | c | c | }

& $h_2 \rightarrow h_1 h_1$& $t\bar t$ & $Z\, b\bar b$& $Z\, jj$  \\
\hline
& $b{\bar b}\tau_\mathrm{had}\tau_\mathrm{had}$& $b{\bar b}
\tau_\mathrm{had}\tau_\mathrm{had}$& $b{\bar b} \tau_\mathrm{had}\tau_\mathrm{had}$ & $jj \tau_\mathrm{had}\tau_\mathrm{had}$  \\
\hline
Event selection (see section V.D)& 6.71  & 509 & 411 &  67 \cr
$\Delta R_{bb} < 2.2$, $P_{T,b_1} > 50$ GeV, $P_{T,b_2} > 30$ GeV& 3.77 & 149 &  58 & 1.68  \cr
$h_1$-mass: $90$ GeV $< m_{bb} < 140$ GeV& 2.99  & 63 & 21.6 & 0.57 \cr
$|\vec{P}^{bb}_T| > 110$ GeV& 2.14 & 32.5 & 9.5  & 0.23 \cr
Collinear $x_1,\,x_2$ Cuts& 1.27 & 13.9 & 4.95 & 0.13  \cr
$\Delta R_{\tau\tau} < 2.3$& 0.92  & 8.1 & 2.51  & 0.034 \cr
$h_1$-mass: $110$ GeV $< m^{\mathrm{coll}}_{\tau\tau} < 150$ GeV& 0.64 & 1.91 & 0.26 & 0.002  \cr
$25$ GeV $ < E_T^\mathrm{miss} < 90$ GeV & 0.47 & 0.98 & 0.19 & 0.001  \cr
$h_2$-mass: $ 330$ GeV $< m^{\mathrm{coll}}_{bb\tau\tau} < 400$ GeV& 0.39  & 0.23 & 0.03  &  $<$ 0.001 \cr
\hline
\end{tabular}
\caption{
Event selection and background reduction for the $b{\bar b}\tau_\mathrm{had}\tau_\mathrm{had}$ channel in the boosted benchmark scenario 
(same efficiency and face rate assumptions as in Table \ref{tab:leplepcutflow1}).}
\label{tab:hadhadcuts_bis}
\end{table*}

\section{Discussion and Outlook}
\label{sec:discuss}

Uncovering the full structure of the SM scalar sector and its possible extensions will be a central task for the LHC in the coming years. 
The results will have important implications not only for our understanding of the mechanism of electroweak symmetry-breaking but also for the origin 
of visible matter and the nature of dark matter. Extensions of the SM scalar sector that address one or both of these open questions may yield distinctive 
signatures at the LHC associated with either modifications of the SM Higgs boson properties and/or the existence of new states. 

In this study, we have considered one class of Higgs portal scalar sector extensions containing a singlet scalar that can mix with the neutral component of 
the SU(2)$_L$ doublet leading to two neutral states $h_{1,2}$. This xSM scenario can give rise to a strong first order electroweak phase transition as needed for 
electroweak baryogenesis; it maps direction onto the NMSSM in the decoupling limit; and it serves as a simple paradigm for mixed state signatures in Higgs 
portal scenarios that contain other SU(2)$_L$ representations. Considering resonant di-Higgs production $pp\to h_2\to h_1 h_1$, we have shown that a search for 
the $b{\bar b}\tau^+\tau^-$ final state could lead to discovery of this scenario with $\sim 100$ fb$^{-1}$ integrated luminosity for regions of the model parameter 
space of interest to cosmology. The most promising mode appears to involve one leptonically-decay and one hadronically-decaying $\tau$ lepton, though for $m_2$ close 
to $2 m_1$ the purely leptonic decay modes of the $\tau$'s 
could also yield discovery as well. For purely hadronically-decay $\tau$ leptons, the significance obtained from our analysis looks promising, though a more refined 
study of the rate for jets faking hadronically decaying $\tau$'s would give one more confidence in the prospects for this mode.

The study of other final states formed from combinations of SM Higgs decay products, as suggested by the  work of Ref.~\cite{Liu:2013woa} that appeared as we 
were completing this paper, would be a natural next step. Although we disagree with the quantitative results in that study (a preliminary application of their basic 
cuts to the $b{\bar b}\tau^+\tau^-$ final state yields $S/B\sim 1$ rather than the $\sim 200$ as these authors find), we concur that a detailed analysis of other 
novel states associated with resonant di-Higgs production would be a worthwhile effort.

\begin{center}
\textbf{Acknowledgements} 
\end{center}

J.M.N. thanks Veronica Sanz for very useful discussions. MJRM thanks B. Brau, C. Dallapiccola,  and S. Willocq  for helpful discussions and 
comments on the manuscript. Both authors thank H. Guo, T. Peng, and H. Patel for generating
background event samples. 
J.M.N. is supported by the Science Technology and Facilities Council (STFC) under grant No.\ ST/J000477/1. MJRM was supported in part
by U.S. Department of Energy contract DE-FG02-08ER41531 and the Wisconsin Alumni Research Foundation. The authors also
thank the Excellence Cluster Universe at the Technical University of Munich, where a portion of this work was carried out. 

\bibliographystyle{h-physrev3.bst}
\bibliography{Singlet_Pheno_MRM_JM_Submission.bib}


\end{document}